\documentclass[iop]{emulateapj}
\usepackage{natbib}

\usepackage{graphicx}

\begin{document}

\title{On the thermal and double episode emissions in GRB 970828}
\author{L. Izzo$^{1,2}$, R. Ruffini$^{1,2,3}$, C. L. Bianco$^{1,2}$, H. Dereli$^{3,4,5}$, M. Muccino$^{1}$, A. V. Penacchioni$^{1,3,5}$, G. Pisani$^{1,3,5}$, Jorge A. Rueda$^{1,2}$}
 
\altaffiltext{1}{Dip. di Fisica and ICRA, Sapienza Universit\`a di Roma, Piazzale Aldo Moro 5, I-00185 Rome, Italy.}
\altaffiltext{2}{ICRANet, Piazza della Repubblica 10, I-65122 Pescara, Italy.}
\altaffiltext{3}{Universite de Nice Sophia Antipolis, CEDEX 2, Grand Chateau Parc Valrose, Nice, France.}
\altaffiltext{4}{I.N.A.F. Osservatorio Astronomico di Capodimonte, Salita Moiariello, 16, 80131, Napoli, Italy.}
\altaffiltext{5}{Erasmus Mundus Joint Doctorate IRAP PhD Student.}

\shorttitle{GRB 970828}

\shortauthors{Izzo et al.}

\newcommand{\Nature}{\mbox{\it Nature}}
\newcommand{\et}{\mbox{\it et al.}}
\newcommand{\ApJ}{\mbox{\it Astrophys. J.}}
\newcommand{\Apjl}{\mbox{\it Astrophys. J.}}
\newcommand{\Aap}{\mbox{\it Astron. Astrophys.}}
\newcommand{\AJ}{\mbox{\it Astron. J.}}
\newcommand{\MNRAS}{\mbox{\it Mon. Not. R. Astron. Soc.}}
\newcommand{\GCN}{\mbox{\it GCN Circ.}}

\begin{abstract}
Following the recent theoretical interpretation of GRB 090618 and GRB 101023, we here interpret GRB 970828 in terms of a double episode emission: the first episode, observed in the first 40 s of the emission, is interpreted as the proto-black-hole emission; the second episode, observed after t$_0$+50 s, as a canonical gamma ray burst. The transition between the two episodes marks the black hole formation. The characteristics of the real GRB, in the second episode, are an energy of $E_{tot}^{e^+e^-} = 1.60 \times 10^{53}$ erg, a baryon load of $B = 7 \times 10^{-3}$ and a bulk Lorentz factor at transparency of $\Gamma = 142.5$. The clear analogy with GRB 090618 would require also in GRB 970828 the presence of a possible supernova. We also infer that the GRB exploded in an environment with a large average particle density $\langle n \rangle \, \approx 10^3$ part/cm$^3$ and dense clouds characterized by typical dimensions of $(4 - 8) \times 10^{14}$ cm and $\delta n/n \propto 10$. Such an environment is in line with the observed large column density absorption, which might have darkened both the supernova emission and the GRB optical afterglow.
\end{abstract}


\maketitle

\section{Introduction}\label{sec:1}

GRB 970828 is one of the first GRBs with an observed afterglow \citep{Piro1999} and a determined redshift of $z$=0.9578 from the identification of its host galaxy \citep{Djorgovski2001}. This source is still presenting today, after 15 years from its observations, an extremely rich problematic in the identification of its astrophysical nature.

The X-ray afterglow was discovered by the ASCA satellite 1.17 days after the GRB trigger \citep{Murakami1997} and the data showed the presence of a feature around $\sim$ 5 keV, maybe associated to a radiative recombination edge of a H- or He-like ionized iron \citep{Yoshida2001}. Particularly interesting was also a variable and large intrinsic absorption column which was found from the ASCA: the absence of an optical afterglow \citep{Groot1998}, the large intrinsic absorption column detected in the X-ray data \citep{Yoshida2001} and the contemporary detection in radio-wavelengths of the GRB afterglow, imply a very large value for the circum-burst medium (CBM), this variable absorption might be an indication of a strong inhomogeneous CBM distribution. 

\citet{Peer2007} described the presence of an evolving thermal component in the first 40 s of the emission of GRB 970828, as observed by the BATSE detector on board the Compton Gamma Ray Telescope \citep{Meegan1992}. This thermal emission was associated to the photospheric GRB emission of a relativistic expanding fireball \citep{Meszaros2002}, and from the observed properties of the blackbody spectrum it was inferred the bulk Lorentz Gamma factor of the expanding plasma, $\Gamma \approx (305 \pm 28)$. This first thermal component is followed by a second emission episode, which starts $\sim$ 50 s after the GRB trigger. This second episode was neglected in the analysis of \citet{Peer2007}, where they explicitly stated to ``neglect here late-time episodes of engine activity that occur after $\sim$ 25 and $\sim$ 60 s in this burst''. 

We have recently proposed that some GRBs, namely GRB 090618 \citep{Ruffini2011,Izzo2012} and GRB 101023 \citep{Penacchioni2012}, can be described by two sharply different emission episodes. A first episode that is characterized by a thermal component in the early phases of its emission, which originates from an expanding source, not related to the GRB emission.
We have shown that this component is characterized by an expansion of the blackbody emitter, whose origin is associated to the latest phase of the life of a massive star. In the late phase of its evolution, this massive star is well-described by a bare core that has lost its Hydrogen envelope and most or entirely the Helium one, during the previous stages of its evolution. This core is then observed to expand at non-relativistic velocities. At the end of the first episode, the collapse to a black hole occurs and the actual GRB emission happens. Since this first episode occurs prior to the GRB emission and to the formation of a black hole, we have referred to this as the proto-black-hole ( PBH - $\pi \rho \tilde{\omega} \tau o \varsigma$ from the ancient greek that means anteriority in space and in time ).
 
In this work we describe first the PBH component that we find in the first episode of GRB 970828 light curve. We then proceed to the description of the second episode, which in this work represents the authentic GRB emission. The GRB emission is well explained in the context of the fireshell scenario, see e.g. \citet{Ruffini2007b} for a complete review of the model. We find for the GRB an $e^+e^-$-plasma total energy of $E_{tot}^{e^+e^-} = 1.60 \times 10^{53}$ erg, a baryon load $B = 7 \times 10^{-3}$ and a Lorentz Gamma factor at transparency of $\Gamma = 142.5$.
From the numerical simulations we confirm the presence in the CBM of large average particle density, of the order of $\langle n \rangle \, \approx 10^3$ part/cm$^3$, and dense clouds with $\delta n/n \propto 10$. We show  that this can explain the ``dark'' nature of this GRB and the observed large and variable column absorption in the ASCA X-ray afterglow.
These results are in very good agreement with the analysis presented in \citet{Yoshida2001} and represents a further confirmation of the presence of a PBH in the early emission of GRB 970828.

\section{Data Analysis}\label{sec:2}

\begin{figure}
\centering      
\includegraphics[width=8cm]{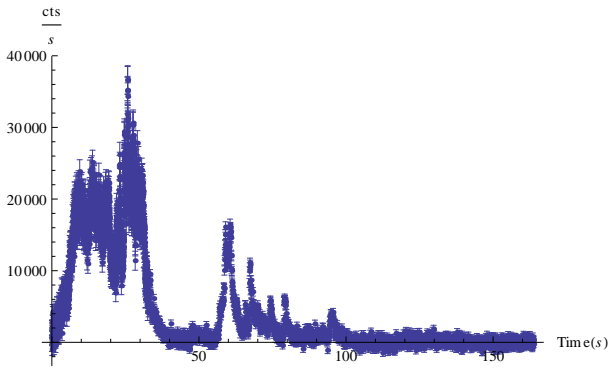}
\caption{BATSE-LAD light curve of GRB 970828 in the 25-1900 keV energy range.}
\label{fig:4b.0}
\end{figure}

GRB 970828 was discovered with the All-Sky Monitor (ASM) on board the Rossi X-Ray Timing Explorer (RXTE) on 1997 August 28th \citep{IAUC6728}.
Within 3.6 hr the RXTE/PCA scanned the region of the sky around the error box of the ASM burst and detected a weak X-ray source \citep{IAUC6727,IAUC6729}.
GRB 970828 was also observed by the Burst and Transient Source Experiment (BATSE) and the GRB experiment on Ulysses \citep{IAUC6728}. 
The BATSE-LAD light curve is characterized by two main emission phenomena, see Fig.\ref{fig:4b.0}: the first lasts about 40 s and is well described by two main pulses, the second one is more irregular, being composed by several sharp pulses, lasting other 40 s.
The X-ray afterglow observations started at about 0.1 days after the trigger time and continued up to 7-10 days.
It is well described by a double power-law behavior, but a possible resurgence of soft X-ray emission, together with marginal evidence for the existence of Fe lines, was reported by Beppo-SAX and ASCA \citep{Vietri1999}.

The optical observations, which started about 4 hr after the burst, did not report any possible optical afterglow for GRB 970828 up to $R$=23.8 \citep{Groot1998}. However, the observations at radio wavelengths of the burst position, 3.5 hrs after the initial burst, succeeded in identifying a source at a good significance level of 4.5 $\sigma$ \citep{Djorgovski2001} inside the ROSAT error circle (10''). The following deep searches for a possible optical counterpart of this radio source led to the identification of an interacting system of faint galaxies, successively recognized as the host galaxy of GRB 970828. The spectroscopic observations of the brightest of this system of galaxies led to the identification of their redshift, being $z$=0.9578. The lack of an optical transient associated with the afterglow of GRB 970828 can be explained as due to the presence of strong absorption, due to dusty clouds in the burst site environment, whose presence does not affect the X-ray and the radio observations of the GRB afterglow.

To analyze in detail this GRB, we have considered the observations of the BATSE-LAD detector, which observed GRB 970828 in the 25-1900 keV energy range, and then we have reduced the data by using the RMFIT software package.
For the spectral analysis we have considered the High Energy Resolution Burst (HERB) data, which consist of 128 separate high energy resolution spectra stored during the burst emission.
The light curve was obtained by using the Medium Energy Resolution (MER) data, which consist of 4.096 16-channel spectra summed from triggered detectors. 

\subsection{The PBH emission in the first emission episode}

\begin{table*}
\centering
\caption{Spectral analysis (25 keV - 1.94 MeV) of the first 40 s of emission in GRB 970828.} 
\vspace{5mm}
\label{tab:4b.1} 
\begin{tabular}{l c c c c c c c }
\hline\hline
Spectral model & $\alpha$ & $\beta$ & $E_0(keV)$ & $kT_1(keV)$ & $kT_2(keV)$ & $\gamma$ & $\tilde{\chi}^2 $\\ 
\hline 
Band & -0.62 $\pm$ 0.03 & -2.14 $\pm$ 0.05 & 358.0 $\pm$ 12.2  &  &  &  & 1.21 \\
BB + po &   &  &  & 63.00 $\pm$ 0.89 & & 1.51 $\pm$ 0.01 & 2.10\\
DoubleBB + po &  &  &  & 108.0 $\pm$ 6.5 & 40.6 $\pm$ 2.0 & 1.54 $\pm$ 0.02 & 0.99\\
\hline
\end{tabular}
\end{table*}

\begin{figure*}
\centering
\begin{tabular}{|c|c|c|}
\hline
\includegraphics[height=4.2cm,width=5.3cm,angle=0]{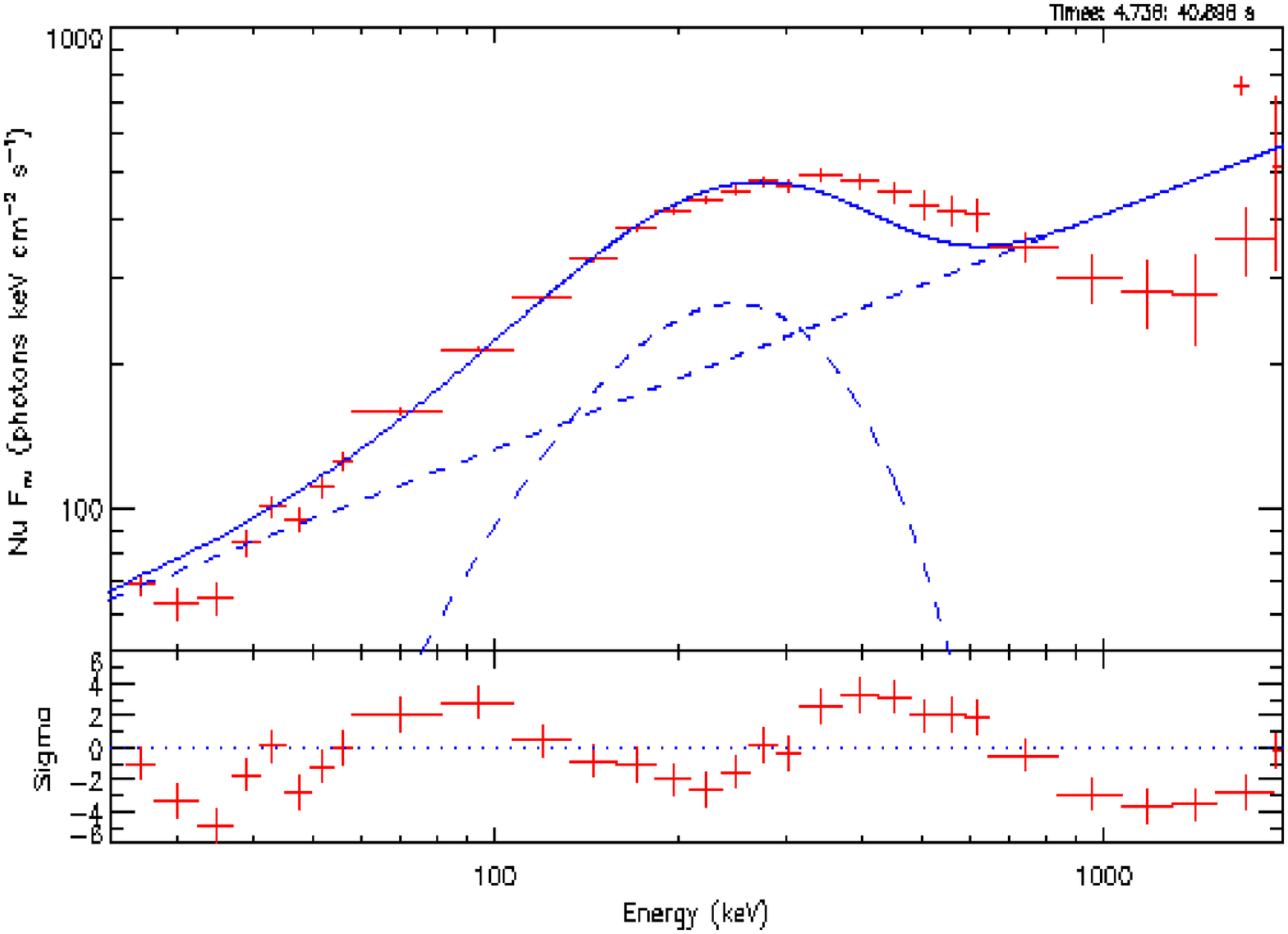}
\includegraphics[height=4.2cm,width=5.3cm,angle=0]{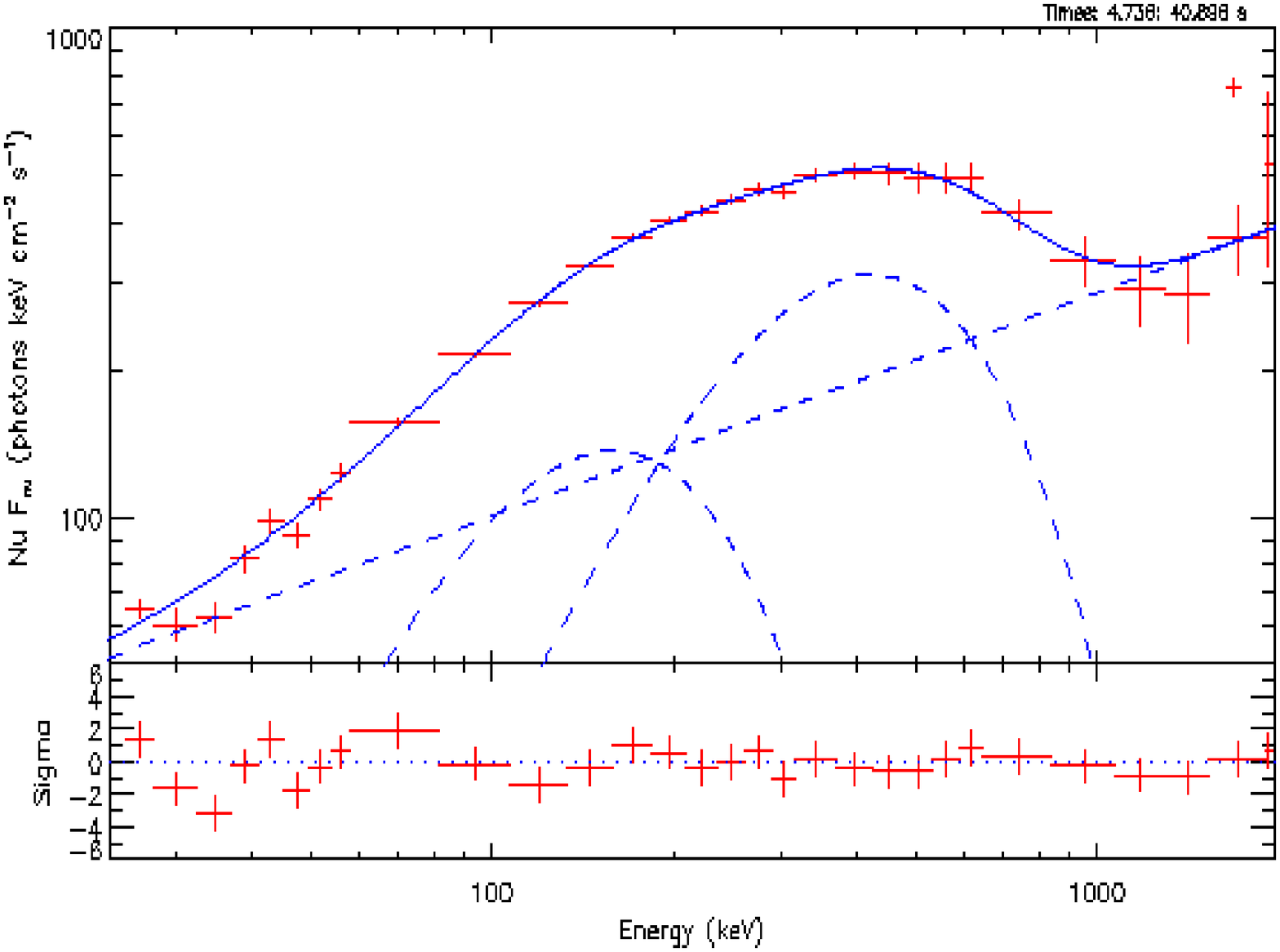}
\includegraphics[height=4.2cm,width=5.3cm,angle=0]{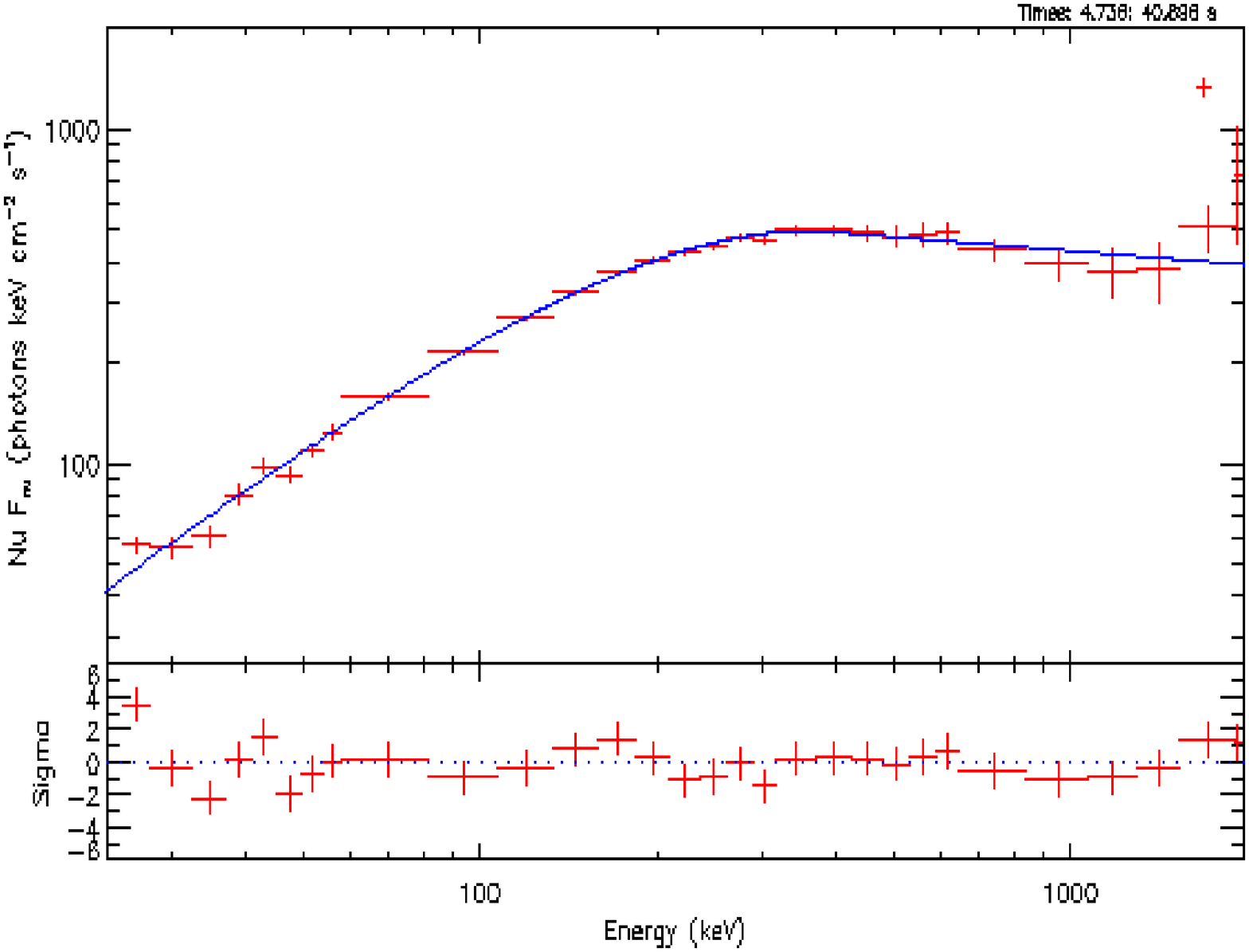}\\
\hline
\end{tabular}
\caption{Time-integrated spectral fits and sigma residual plot (25-1900 keV) of the first emission episode (0-40 s) in GRB 970828 with respectively a) a blackbody plus an extra power-law model; b) a double blackbody plus an extra power-law component; c) a Band model.}
\label{fig:4b.boh}
\end{figure*}

In analogy to the cases of GRB 090618 \citep{Izzo2012} and GRB 101023 \citep{Penacchioni2012}, we analyze here the first emission episode in GRB 970828 to seek for a PBH signature. An integrated spectral analysis of these initial 40 s of emission is reported in Table \ref{tab:4b.1}.
This first time interval is best fitted by a Band spectral model and by a double blackbody with an extra power-law component.
A single blackbody with power-law is not satisfactory, providing a $\tilde{\chi}^2 = 2.10$, see Fig. \ref{fig:4b.boh}.

\begin{figure}
\centering      
\includegraphics[width=9cm]{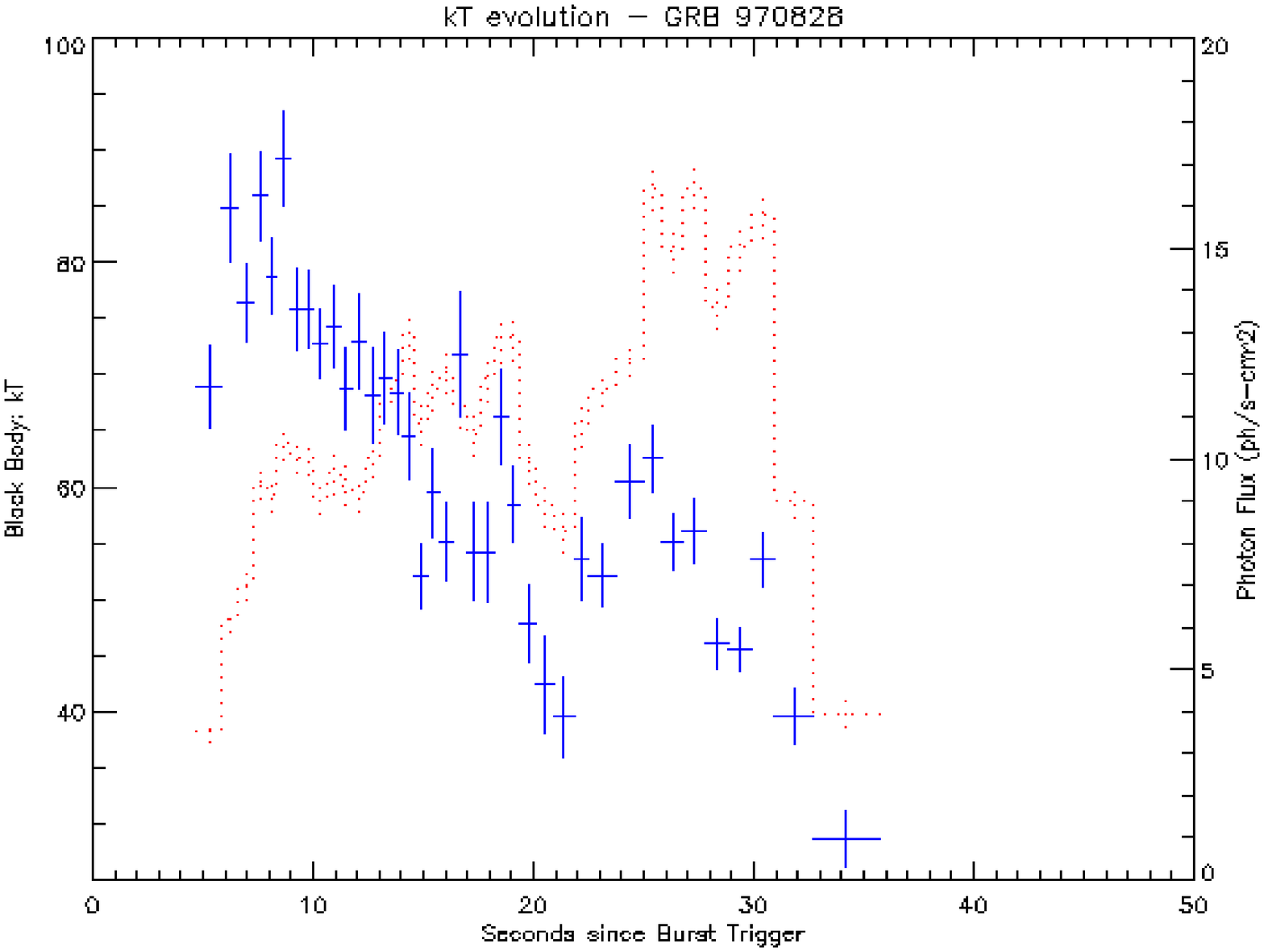}
\caption{The evolution of the temperature $kT$ (blue bars) as obtained from a time-resolved spectral analysis of the first 40 s of emission of GRB 970828. The light curve of the first episode (red dots) is shown in background.}
\label{fig:4b.1}
\end{figure}

\begin{figure}
\centering
\includegraphics[width=8cm, height=5cm, angle=0]{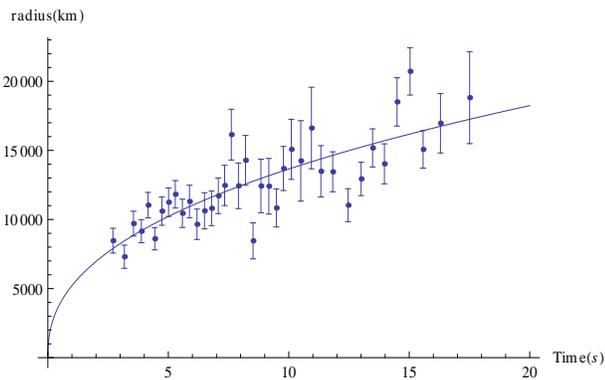}
\caption{Evolution of the radius of the first episode of GRB 970828. The line corresponds to the best fit of this dataset with a power-law function r $\propto$ t$^{\delta}$, with $\delta = 0.42 \pm 0.05$. }
\label{fig:4b.2}
\end{figure}
 
A detailed time-resolved investigation of these initial 40 s of emission has provided very interesting results: the analysis with a single blackbody plus a power-law model has revealed a double decay trend  of the temperature $kT$, corresponding to the two main spikes in the observed light curve of this first episode, see Fig. \ref{fig:4b.1}.
We have then analyzed this characteristic evolution of the blackbody in both time intervals, corresponding each one to an observed decay trend of the temperature.
From the observed flux of the blackbody component $\phi_{BB,obs}$ for each interval, we obtain the evolution of the emitter radius, which is given by
\begin{equation}
r_{em} = \left(\frac{\phi_{BB,obs}}{\sigma T^4}\right)^{1/2} \frac{D}{(1+z)^2},
\end{equation}
where $\sigma$ is the Stefan constant, $T$ the observed temperature, $D$ the luminosity distance of the GRB and $z$ the redshift.
The evolution of the emitter radius for the complete PBH emission is shown in Fig. \ref{fig:4b.2}. 
It is very interesting that the radius monotonically increases, without showing the double trend observed for the temperature, see Fig. \ref{fig:4b.1}.
The global evolution of the emitter radius is well-described with a power-law function $r \propto t^{\delta}$ and a best fit of the data provides for the $\delta$ index the value of 0.42 $\pm$ 0.05, with an $R^2$ statistic value of 0.98, see Fig. \ref{fig:4b.2}.

We conclude that the double blackbody observed in the integrated emission of the first episode is indeed due to the expansion, during the PBH emission, from an initial radius of 7000 km to a final distances of $\sim$ 20000 km from the progenitor.

\subsection{The GRB emission in the second emission episode}

The observed fluence in GRB 970828 and the redshift $z = 0.9578$ imply an isotropic energy for the total GRB emission $E_{iso}=4.2 \times 10^{53}$ erg.
Following the separation of this GRB in two different episodes we have computed the isotropic energies emitted in both episodes, by considering a Band model as the best fit for the observed integrated spectra: $E_{iso,1st} = 2.6 \times 10^{53}$ erg and $E_{iso,2nd} = 1.6 \times 10^{53}$ erg.
In what follows we explain the second emission episode of GRB 970828 as a single canonical GRB emission in the context of the Fireshell scenario.

In this model \citep{Ruffini2001L107,Ruffini2001K,Ruffini2001c}, a GRB originates from an optically thick $e^+e^-$-plasma created in the process of vacuum polarization by a charged black hole \citep{Christodoulou,Damour}. The dynamics of this expanding plasma is mainly described by its total energy $E_{tot}^{e^+e^-}$, the baryon load $B = M_B c^2/E_{tot}^{e^+e^-}$ and the circum burst matter (CBM) distribution around the burst site. The GRB light curve emission is characterized by a first brief emission, named the proper GRB or P-GRB, originating in the process of the transparency emission of the $e^+e^-$-plasma, followed by a multi-wavelength emission due to the collisions of the residual accelerated baryons and leptons with the CBM. This latter emission is assumed in a fully radiative regime and is called the extended afterglow. In the spherically symmetric approximation
 the interaction of the accelerated plasma with the CBM can be described by the matter density distribution $n_{CBM}$ around the burst site and the fireshell surface filling factor $\mathcal{R} = A_{eff}/A_{vis}$, which is the ratio between the effective emitting area and the total one \citep{Ruffini2005}. The spectral energy distribution is well-described by a ``modified'' thermal emission model \citep{Patricelli2011}, which differs from a classical blackbody model by the presence of a tail in the low-energy range.

\begin{figure*}
\centering
\begin{tabular}{|c|c|}
\hline
\includegraphics[height=6cm,width=8cm,angle=0]{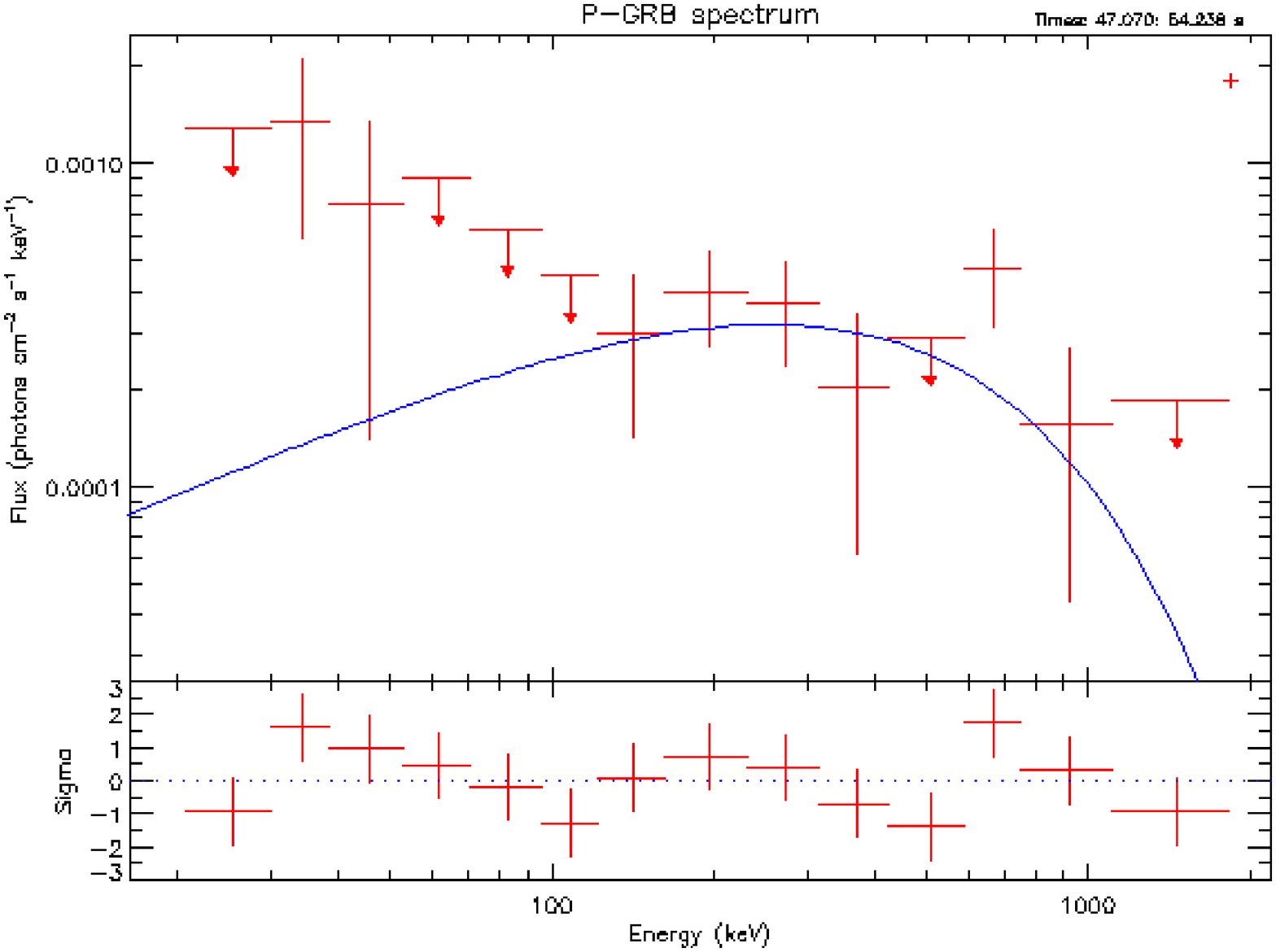}
\includegraphics[height=6cm,width=8cm,angle=0]{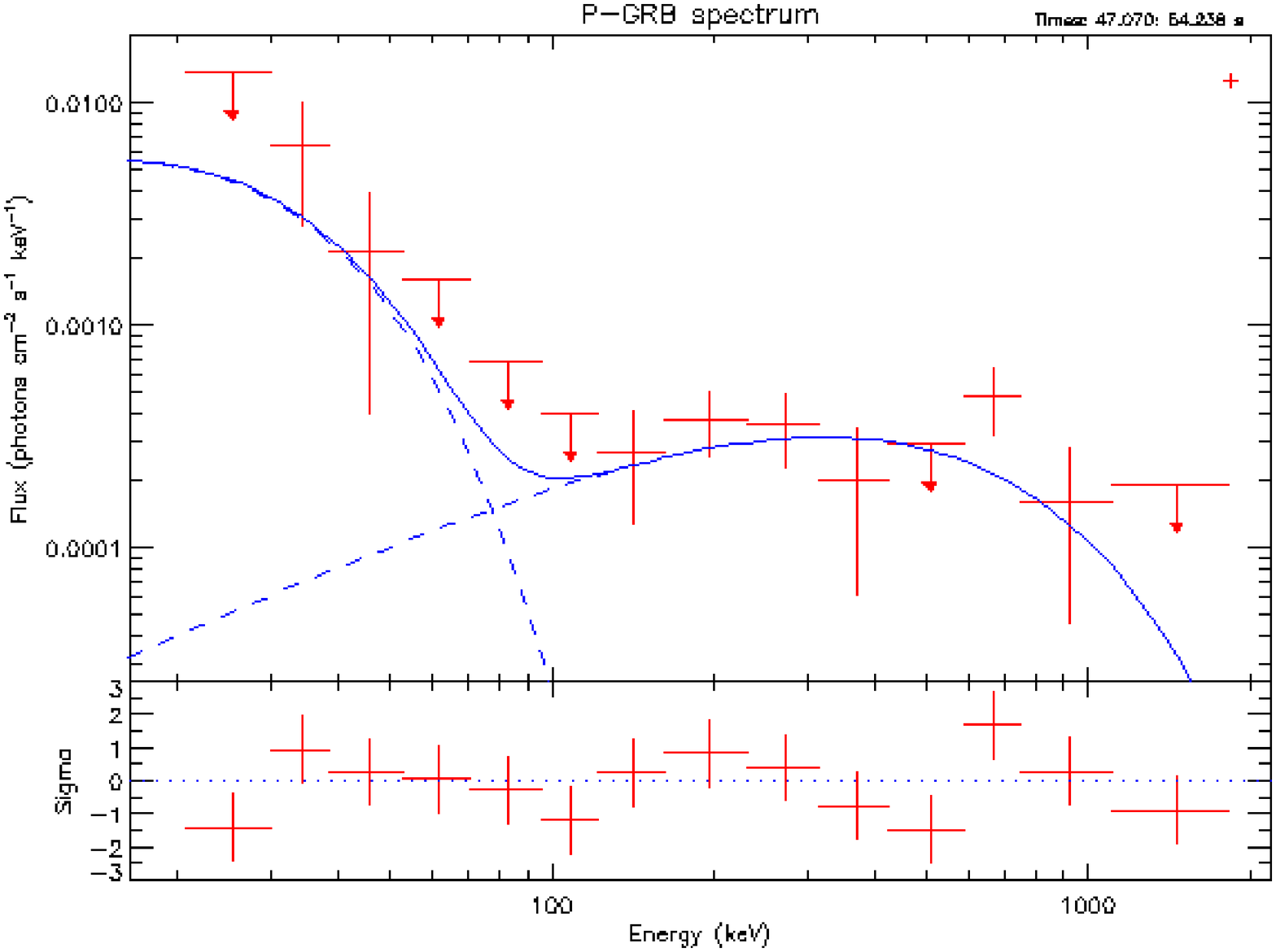}\\
\hline
\end{tabular}
\caption{Time-integrated spectral fits and sigma residual plot (25-1900 keV) of the P-GRB emission in the second episode (0-40 s) in GRB 970828 with respectively a) a power-law model with an exponential cut-off; b) a blackbody plus a power-law model with an exponential cut-off.}
\label{fig:4b.boh2}
\end{figure*}

\begin{figure}
\centering
\includegraphics[width=9cm, height=7cm, angle=0]{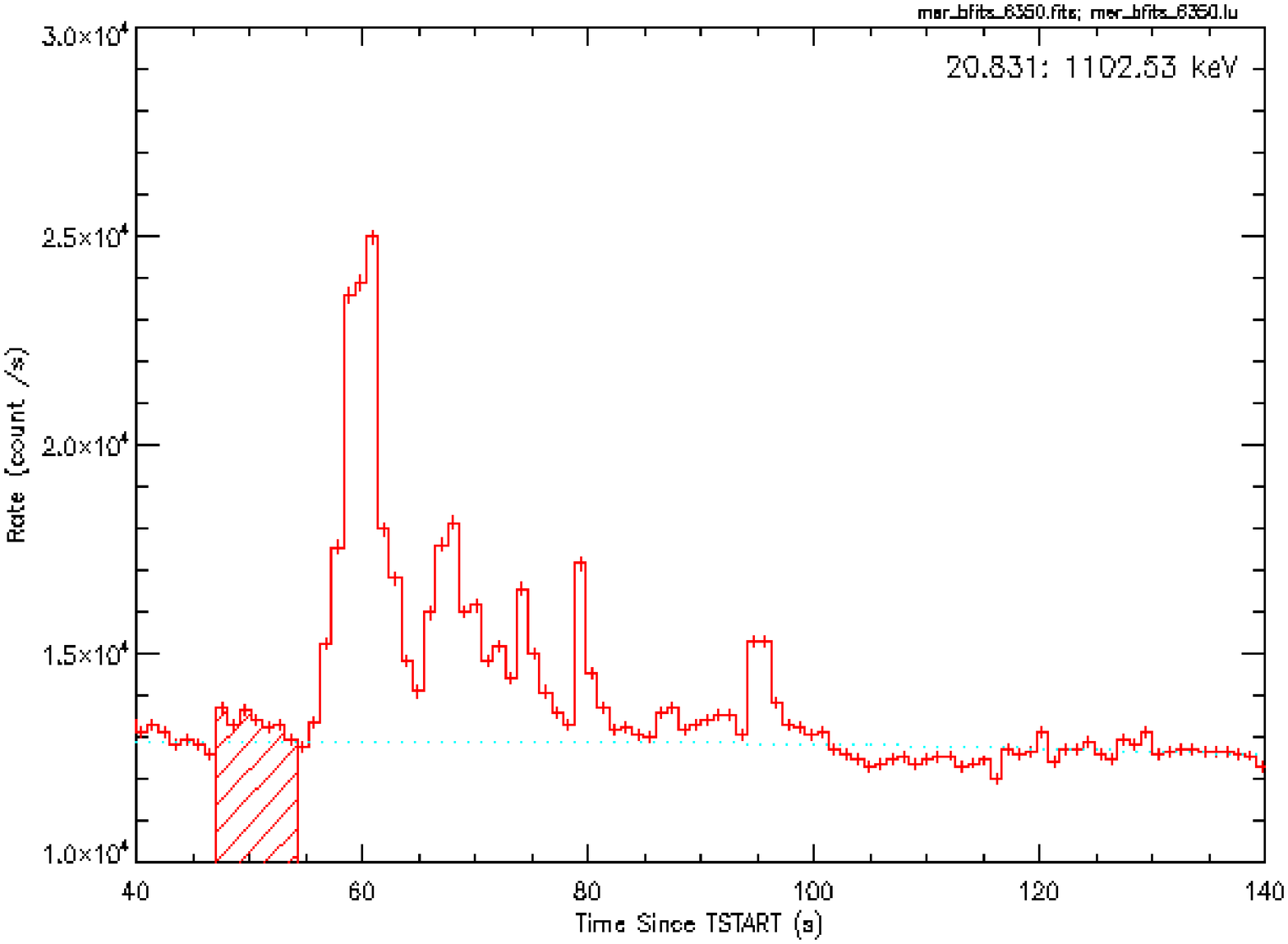}\\
\caption{Light curve of the second episode in GRB 970828. The dashed region represents the P-GRB emission. }
\label{fig:4b.3b}
\end{figure}

In this context, to simulate the second episode of GRB 970828, which is the actual GRB emission, we need to identify the P-GRB signature in the early second episode light curve.
From the identification of the P-GRB thermal signature, and the consequent determination of the energy emitted at transparency, we can obtain the value of the baryon load $B$ assuming that the total energy of the $e^+e^-$-plasma is given by the isotropic energy $E_{iso}$ observed for the second episode of GRB 970828, as it was done for the second episode in GRB 090618, see e.g. \citet{Izzo2012}.
We have then started  to seek for any thermal feature attributable to the P-GRB emission in the early emission of the second episode.
As it is shown in Fig. \ref{fig:4b.3b}, the early emission of the second episode is characterized by an intense spike, anticipated by a weak emission of 7 s.
The emission observed in this time interval can be fitted by a power-law with an exponential cut-off model (Comptonization model in RMFIT) as well as with the addition of a blackbody component, see Table \ref{tab:4b.2} and Fig. \ref{fig:4b.boh2}.

\begin{table*}
\centering
\caption{Spectral analysis (25 keV - 1.94 MeV) of the P-GRB emission in the second episode of GRB 970828.} 
\vspace{5mm}
\label{tab:4b.2} 
\begin{tabular}{l c c c c}
\hline\hline
Spectral model & $\gamma$ & $E_{cutoff}(keV)$ & $kT(keV)$ & C-STAT/dof  \\ 
\hline 
Compt & -0.68 $\pm$ 0.42 & 969.8 $\pm$ 180  &  & 75.3/11 \\
BB + Compt & -1.13 $\pm$ 0.58 & 894.4 $\pm$ 152 & 9.23 $\pm$ 2.46 & 66.7/9\\
\hline
\end{tabular}
\end{table*}

\begin{figure}
\centering
\includegraphics[width=9cm, height=7cm, angle=0]{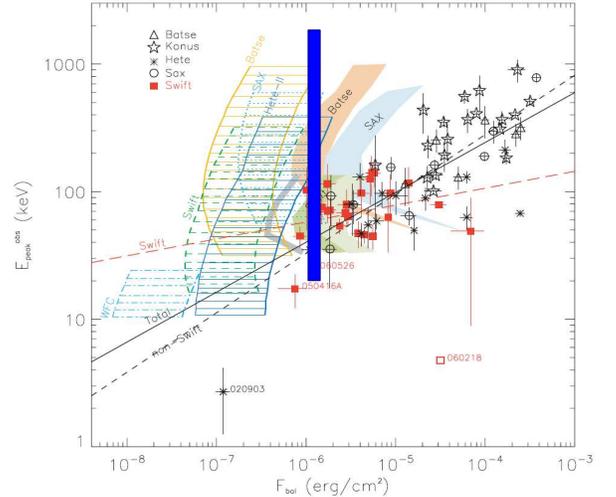}
\caption{Observer frame plane of the peak energy and bolometric fluence (1 - 10$^4$ keV). The blue region represents the value of the observed fluence for the 7 s of emission of the second episode in GRB 970828, where the P-GRB emission is identified. The shaded orange region represent the minimum fluence that a burst detected by BATSE should have in order to perform the spectral analysis and make constraints about the main spectral parameters. The figure is taken from the work of \citet{Ghirlanda2008}. }
\label{fig:4b.3}
\end{figure}

The observed fluence in the P-GRB emission, computed from the fit with the blackbody and the power-law with exponential cut-off model is $S_{obs} = 1.16 \times 10^{-6}$.
This value of the fluence is at the boundary of the region of minimum fluence that a burst should have in order to perform a spectral analysis, in a given energy range, and then derive the main spectral parameters, as the peak energy or the blackbody temperature \citep{Ghirlanda2008}, see also Fig. \ref{fig:4b.3}.
In this light, we can conclude that the signature of the P-GRB emission is represented by these early 7 s of emission and in the following we will base on this time interval for the P-GRB emission of the second episode of GRB 970828.

\begin{table}
\centering
\caption{Final results of the simulation of GRB 970828 in the fireshell scenario} 
\vspace{5mm}
\label{tab:4b.3} 
\begin{tabular}{l c}
\hline\hline
Parameter & Value \\ 
\hline 
$E_{tot}^{e^+e^-}$ &  (1.60 $\pm$ 0.03) $\times$ 10$^{53}$ erg\\
$B$ &  (7.00 $\pm$ 0.55) $\times$ 10$^{-3}$\\
$\Gamma_0$ & 142.5 $\pm$ 57\\
$kT_{th}$ & (7.4 $\pm$ 1.3) keV\\
$E_{P-GRB,th}$ & (1.46 $\pm$ 0.43) $\times$ 10$^{51}$ erg\\ 
$\langle  n  \rangle $ & $3.4 \times 10^3 \, part/cm^3$\\   
$\delta n/n$ & $10 \, part/cm^3$\\
\hline
\end{tabular}
\end{table}

\begin{figure}
\centering
\includegraphics[width=9cm, height=6cm, angle=0]{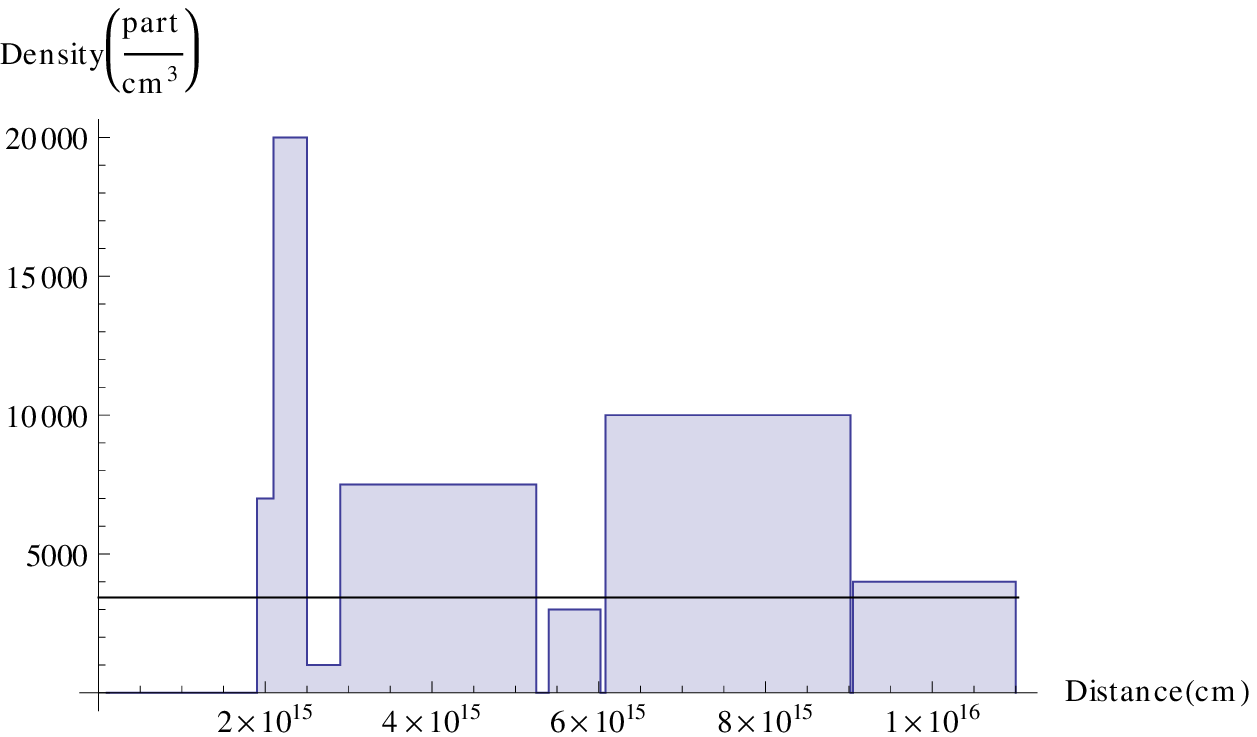}\\
\caption{Radial CBM density distribution for GRB 970828. The characteristic masses of each cloud are on the order of $\sim$ 10$^{22}$ g and 10$^{15}$ cm in radii. The black line corresponds to the average value for the particle density.}
\label{fig:4b.5}
\end{figure}

From the observed fluence and the values obtained from the spectral fits, we have computed the energy emitted in the P-GRB by using the Schaefer formula of the K-correction \citep{Schaefer2007} and the redshift of the source: it is $E_{PGRB}=1.46\times 10^{51}$ erg.
This value of the P-GRB energy is only 9 $\times$ 10$^{-3}$ the isotropic energy of the second episode, which we can assume as equal to the energy $E_{tot}^{e^+e^-}$ of the $e^+e^-$-plasma of the GRB emission.
In this way, with the knowledge of the temperature at transparency and of the P-GRB energy, we can derive the value of the baryon load from the numerical solutions of the fireshell equations of motion.
These solutions for four different values of the total e$^+$e$^-$-plasma energy are shown in the Fig. 4 of \citet{Izzo2012}.
We find that the baryon load is $B = 7 \times 10^{-3}$, which corresponds to a Lorentz gamma factor at transparency $\Gamma = 142.5$.
The GRB emission was simulated with very good approximation by using a density mask characterized by an irregular behavior: all the spikes correspond to spherical clouds with a large particle density $\langle n \rangle  \sim 10^3$ part/cm$^3$, and with radius of the order of $(4 - 8) \times 10^{14}$ cm, see Fig. \ref{fig:4b.5}.
Considering all the clouds found in our analysis, the average density of the CBM medium is $\langle n \rangle  = 3.4 \times 10^3$ particles/cm$^3$.
The corresponding masses of the blobs are of the order of $10^{22}$ g, in agreement with the clumps found in GRB 090618.

\begin{figure}
\centering
\includegraphics[width=6cm, height=8cm, angle=270]{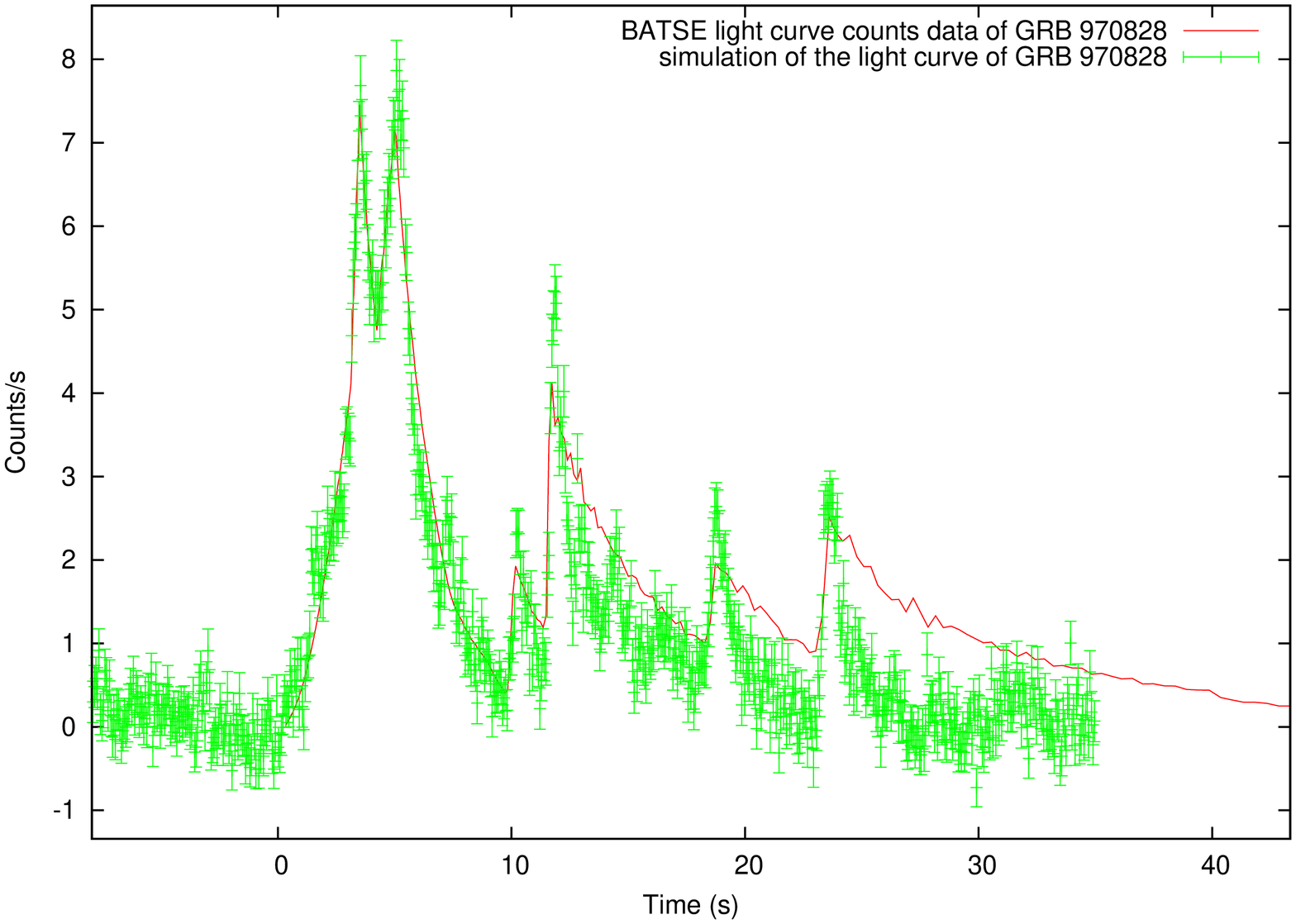}\\
\includegraphics[width=8cm, height=6cm, angle=0]{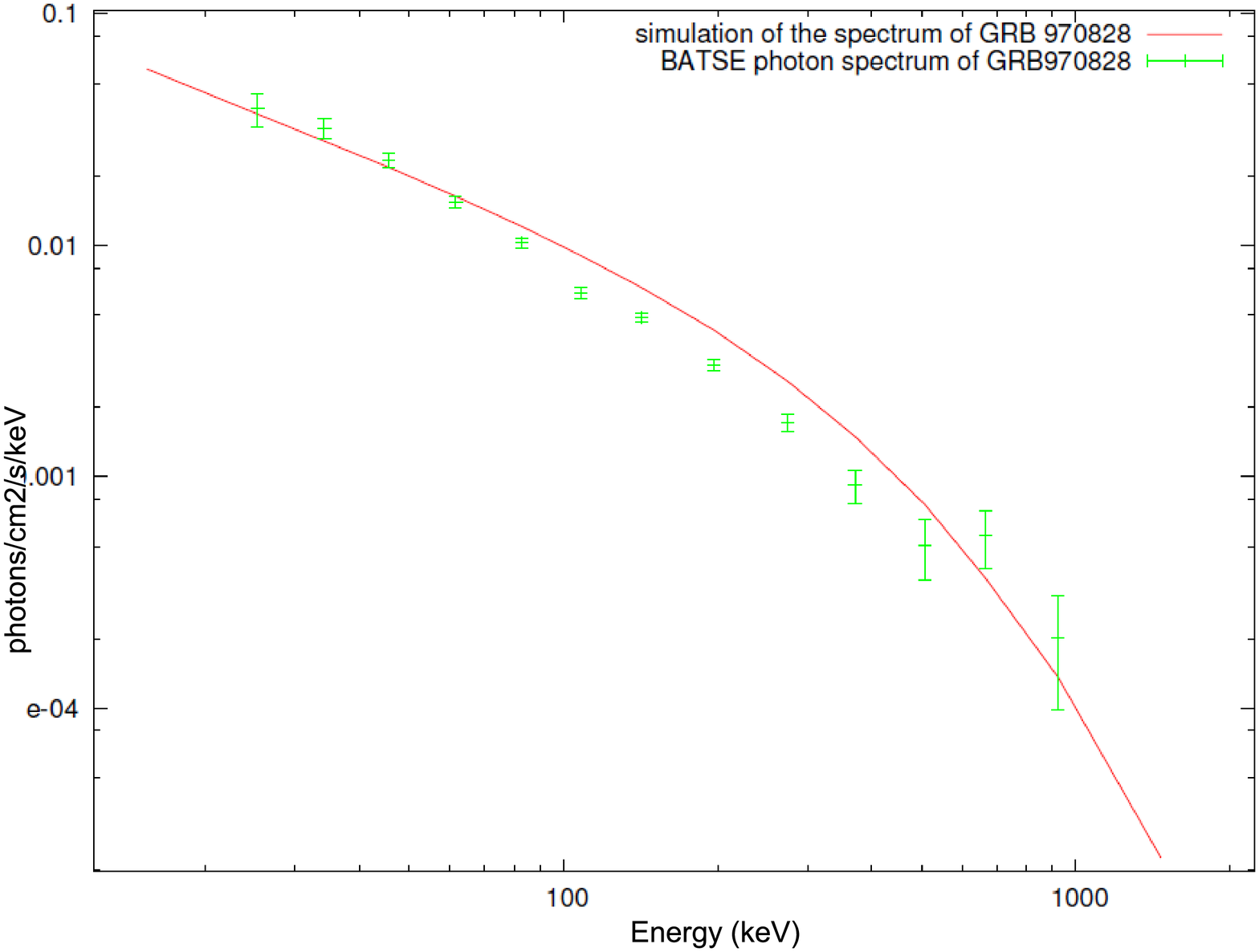}\\
\caption{Light curve with the fireshell simulation (top panel) and spectrum (low panel) with the fireshell simulation of the second episode emission in GRB 970828, in the context of the fireshell scenario. The red lines correspond to the fireshell simulation. }
\label{fig:4b.4}
\end{figure}

The results of the fireshell simulation are shown in Table \ref{tab:4b.3} while the simulated light curve and  spectrum are shown in Fig. \ref{fig:4b.4}.

\begin{figure}
\centering
\includegraphics[width=6cm, height=9cm, angle=270]{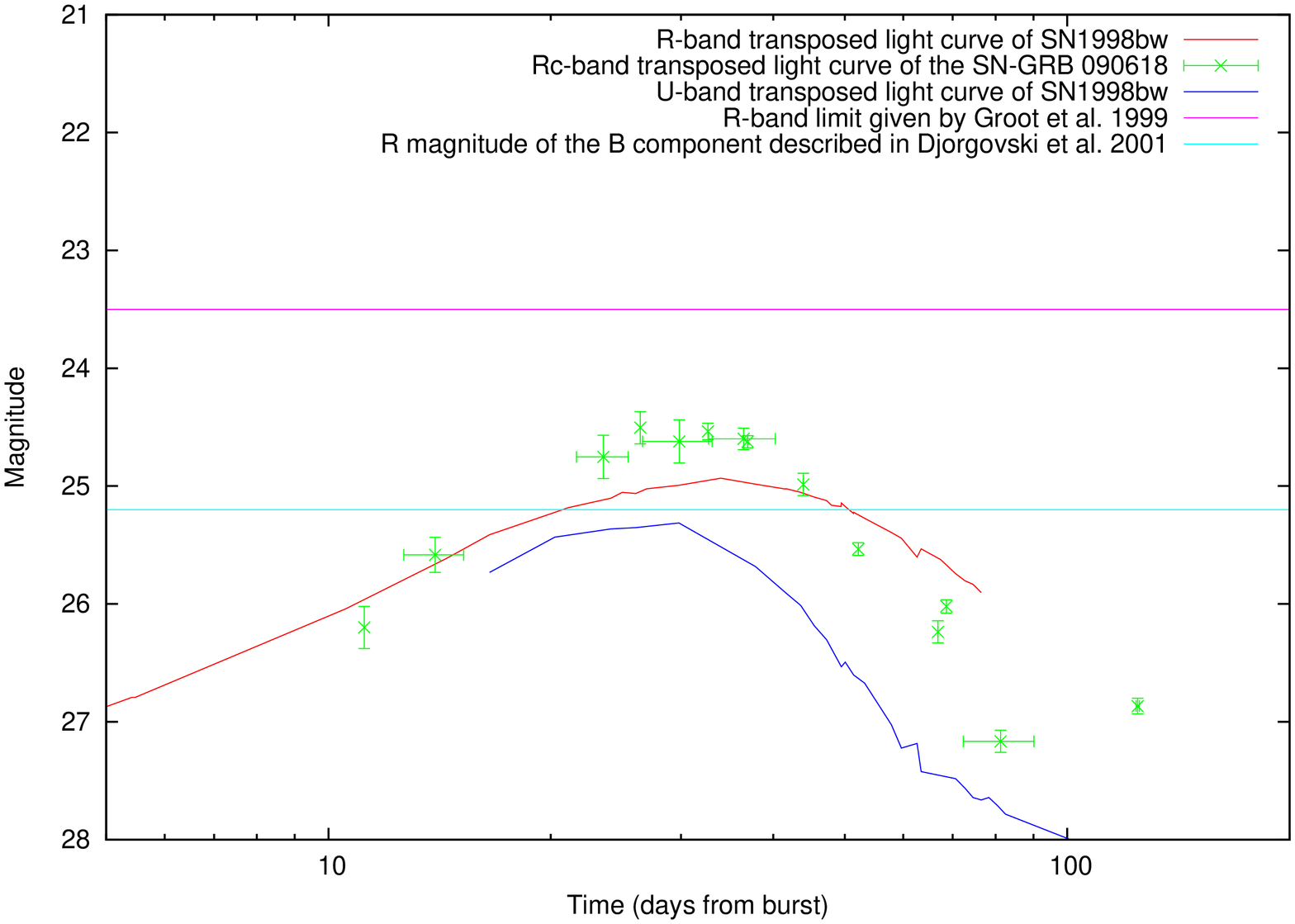}
\caption{The transposed data at the redshift of GRB 970828 of the same SN data (green data) with the $U-$ (blue line) and $R-$band (red line) light curve of SN 1998bw transposed at the same redshift of GRB 970828 . The purple and cyan line represents the limit given in the deep images by \citet{Djorgovski2001} and \citet{Groot1998} respectively.}
\label{fig:4b.6}
\end{figure} 

\subsection{Limits on a possible SN-related observations and neo-neutron star}

The analysis of GRB 090618 \citep{Izzo2012} and GRB 101023 \citep{Penacchioni2012} represents an authentic ``Rosetta Stone'' for the understanding of the GRB-SN phenomenon. The possible presence of a supernova emission, observed ten days after the burst in GRB 090618 \citep{Cano2011} but not spectroscopically confirmed, is at the same time very intriguing in light of the still unknown GRB-SN connection \citep{DellaValle2011}. We have transposed the data of the ``bump'' Rc-band light curve observed in the optical afterglow of GRB 090618, associated to the presence of a possible underlying supernova \citep{Cano2011}, to the redshift of GRB 970828. This simple operation concerns only the transformation of the flux observed, under the assumption that the SN has the same intrinsic luminosity: the variation of the magnitude only depends on the square of the difference between the luminosity distances. Moreover, we have also transposed the $U$ and $R-$band light curves of SN 1998bw \citep{Galama1998}, which is the prototype of a supernova associated to a GRB. From the $K-$ correction transformation formula, the $U-$band light curve, transposed at $z$=0.9578, corresponds approximately to the observed $R-$band light curve, so in principle we should consider the $U = 365$ nm transposed light curve as the actual one observed with the $Rc = 647$ nm optical filter. These transposed light curves are shown in Fig. \ref{fig:4b.6}. We conclude that the possible Supernova emission could have been seen between 20 and 40 days after the GRB trigger, neglecting any possible intrinsic extinction. The optical observations were made up to 7 days from the GRB trigger, reaching a limit of $R \sim 23.8$ \citep{Groot1998} and subsequent deeper images after $\sim 60$ days \citep{Djorgovski2001}, so there are no observations in this time interval. However, the $R$-band extinction value should be large since the observed column density from the X-ray observations of the GRB afterglow is large as well \citep{Yoshida2001}: the computed light curve for the possible SN of GRB 970828 should be lowered by more than 1 magnitude, leading to a SN bump below the $R$ = 25.2 limit, see Fig. \ref{fig:4b.6}. The presence of very dense clouds of matter near the burst site might be the cause of these absorptions. Indeed we find the presence of clouds in our simulation at the average distances of $\sim 10^{15-16}$ cm from the GRB progenitor, with average density of $\langle n \rangle \, \approx 10^3$ part/cm$^3$ and typical dimensions of $(4 - 8) \times 10^{14}$ cm, see Fig. \ref{fig:4b.5}.

The possibility to observe the energy distribution from a GRB in a very wide energy range, thanks to the newly space missions dedicated, has allowed to definitely confirm the presence of two separate emission episodes in GRBs associated to SNe. Future missions, as the proposed LOFT \footnote{http://www.isdc.unige.ch/loft/index.php/the-loft-mission}, will allow to observe the thermal decay from these objects down to $kT = 0.5 -1$ keV. The Large Area Detector, designed for the LOFT mission, could be also able to observe in timing mode the emission from the newly born neutron star (NS) from Supernovae Ib/c, catching a possible characteristic signature of the neo-NS \citep{Negreiros2012}, shedding more light on the SN-GRB enigma.

\section{Conclusions}

A general conclusion is therefore appropriate: it is mandatory to proceed to a detailed spectral analysis in parallel to an identification of the physical and astrophysical nature of the source. This has been made explicitly clear in the analysis of GRB 090618 \citep{Izzo2012}. The crucial role of GRB 090618 has been to evidence the existence of a new family of GRBs, associated to SNe, composed of two distinct emission episodes. 

This has been made possible thanks to different reasons: 1) the large number of detectors, Fermi, Swift BAT and XRT, AGILE, Suzaku-WAM, Konus-WIND and Coronas PHOTON RT-2, that allowed to study that GRB in a very wide electromagnetic energy range, from 0.3 keV to 40 MeV; 2) the very large intensity emitted from GRB 090618, $E_{iso} = 2.7 \times 10^{53}$ erg;  3) the low redshift of  the GRB, $z$ = 0.54.  Although both episodes radiate in the hard X-ray energy range, the first episode originates from a non-relativistic expansion process, having an expansion velocity of $\sim$ 1000 km/s, see Fig. \ref{fig:4b.2}, and occurring prior to the formation of a black hole; the second one is the one occurring at the formation of the black hole and is the authentic GRB emission. 
Of course, the understanding of these two components cannot be done only on a phenomenological basis, but it requires an understanding of the underlying physical and astrophysical processes. 

We verify in this work that indeed GRB 970828 is a member of this family. The new understanding leads to a wealth of information on the evolution of the thermal component and of the radius of the blackbody emitter, given by Eq. (1), in the first emission episode, see Figs. \ref{fig:4b.1}, \ref{fig:4b.2}. Similarly, for the second episode we obtained the CBM structure, see Fig. \ref{fig:4b.5}, the details of the simulation of the lightcurve and the spectrum of the real GRB emission, see Fig. \ref{fig:4b.4}.
We have also shown in Table \ref{tab:4b.2} the final results of the GRB simulation, the total energy of the $e^+e^-$ plasma, the baryon load $B$, the temperature of the P-GRB $kT_{th}$ and the Lorentz Gamma factor at transparency $\Gamma$, as well the average value of the CBM density $\langle n_{CBM} \rangle$ and the density ratio of the clouds $\delta n/n$.
Finally, from the analogy with the optical bump observed in GRB 090618, see Fig. \ref{fig:4b.6}, associated to a possible SN emission \citep{Cano2011}, we have given reasons why a possible SN associated to GRB 970828 would not have been observable due to the large interstellar local absorption, in agreement with the large column density observed in the ASCA X-ray data \cite{Yoshida2001} and with the large value we have inferred for the CBM density distribution, $\langle n_{CBM} \rangle \approx 10^3$ particles/cm$^3$.

\vspace{1cm}

We thank J. Rafelski for useful discussions about the writing of this manuscript. We also thank Z. Cano for providing the data of the optical bump observed in the afterglow of GRB 090618. HD, GP and AVP acknowledge the support for the fellowship awarded for the Erasmus Mundus IRAP PhD program.

\bibliographystyle{apj}

\end{document}